\documentclass[twocolumn,english,aps,prb,floatfix,preprintnumbers,showpacs,amsfonts,amssymb]{revtex4}
\usepackage[T1]{fontenc}
\usepackage[latin1]{inputenc}
\usepackage{amsmath}
\usepackage{amssymb}

\makeatletter

\providecommand{\LyX}{L\kern-.1667em\lower.25em\hbox{Y}\kern-.125emX\@}

\usepackage{graphicx}
\usepackage{graphicx}
\usepackage{graphicx}
\usepackage{amscd}
\usepackage{bm}

\usepackage{babel}
\makeatother
\begin{document}
\title{Transport properties of a quantum wire: the role of extended time-dependent impurities}
\newcommand{\be}{\begin{equation}}
\newcommand{\ee}{\end{equation}}
\newcommand{\bd}{\begin{displaymath}}
\newcommand{\ed}{\end{displaymath}}

\author{D.\ Makogon$^1$, V.\ Juricic$^{1,2}$, and C.\ Morais Smith$^1$}

\affiliation{$^1$Institute for Theoretical Physics, University of
Utrecht, Leuvenlaan 4, 3584 CE Utrecht, The Netherlands.\\
$^2$Department of Physics, Simon Fraser University, Burnaby,
British Columbia, Canada V5A 1S6}

\date{\today{}}

\begin{abstract}
 We study the transport properties of a quantum wire, described by
the Tomonaga-Luttinger model, in the presence of a backscattering
potential provided by several extended time-dependent impurities
(barriers). Employing the B\" uttiker-Landauer approach, we first
consider the scattering of noninteracting electrons ($g=1$) by a
rectangular-like barrier and find an exact solution for the
backscattering current, as well as a perturbative solution for a
weak static potential with an arbitrary shape. We then include
electron-electron interactions and use the Keldysh formalism
combined with the bosonization technique to study oscillating
extended barriers. We show that the backscattering current off
time-dependent impurities can be expressed in terms of the current
for the corresponding static barrier. Then we determine the
backscattering current for a static extended potential, which, in
the limit of noninteracting electrons ($g=1$), coincides with the
result obtained using the B\" uttiker-Landauer formalism. In
particular, we find that the conductance can be increased beyond
its quantized value in the whole range of repulsive interactions
$0<g<1$ already in the case
of a single oscillating extended impurity, in contrast 
to the case of a point-like impurity, where this phenomenon occurs
only for $0<g<1/2$.
\end{abstract}

\pacs{}

\maketitle

\section{Introduction}
The electronic transport in mesoscopic systems has been
intensively studied during the last decades.\cite{Mbook1, Mbook2}
The conductance of mesoscopic wires is usually computed using the
B\"uttiker-Landauer formalism, \cite{BLandauer} which accounts for
the quantum mechanical nature of transmission and reflection
through impurities and works well for Fermi liquids.

Recent advances in nanotechnology, together with the discovery of
carbon nanotubes \cite{SumioIijima, CeesDekker} have enabled the
fabrication of extremely narrow wires, which are basically
one-dimensional. Because the electron-electron interactions become
essential in 1D, these quantum wires exhibit Luttinger liquid
behavior, i.e., they have bosonic collective excitations, instead
of the usual fermionic quasi-particles present in conventional
Fermi liquids.\cite{indianLL}

The role played by impurities in a Luttinger liquid depends
crucially on the nature of the interactions. In pathbreaking
papers, Kane and Fisher\cite{KaneFisher1,KaneFisher2} showed that
the impurity barrier is irrelevant if the electrons have
attractive interactions ($g > 1$), but it cuts the wire into two
pieces if they interact repulsively ($g < 1$). For the
noninteracting case ($g = 1$) the barrier is a marginal
perturbation, i.e., one can have both, transmission and
reflection.

For the case of two barriers, contrarily to the expectation, it is
possible to have transmission even in the presence of repulsive
interactions ($g < 1$) because quasi-bound-states can form between
the barriers, which lead to resonant
transmission.\cite{KaneFisher3}

The vast majority of cases discussed in the literature concerning
transport in  Luttinger liquids involve {\it static} {\it
point-like} impurity barriers. Recently, the case of {\it
dynamical point-like} impurities was addressed.\cite{Sharma,
Schmeltzer, FeldmanGefen, ChengZhou, us} Feldman and
Gefen\cite{FeldmanGefen} have shown that in the presence of one
impurity oscillating with frequency $\Omega$, the backscattering
current may change sign and the conductance of a single-mode
Luttinger liquid with strong repulsive interactions ($g < 1/2$)
may become {\it larger} than the quantum of conductance $G_0 =
e^2/h$. We have generalized their results and shown that in the
presence of several impurities this phenomenon occurs even for
$1/2 < g < 1$.\cite{us} In fact, we have found that the dc-current
can always be expressed in terms of the backscattering current off
a static impurity. The effect of the impurity frequency $\Omega$
is solely to split the energy modes into $\omega_0 + \Omega$ and
$\omega_0 - \Omega$, where $\omega_0 = eV/\hbar$ denotes the
Josephson frequency, associated with the external voltage $V$.
Hence, within first order perturbation theory, several point-like
impurities oscillating in phase may lead to an increase of the
conductance, in a certain range of parameters, but in the entire
regime of repulsive interactions $0 < g < 1$.\cite{us}

After having understood the role of {\it time-dependent point-like
impurities}, the remaining task is to investigate transport
through {\it extended time-dependent} impurities.\cite{Barci} This
is precisely the aim of this paper. The coherent transport
properties of multimode quantum channels in the presence of
Gaussian-like scatterers was considered recently.\cite{Bardarson}
In addition, the tunneling current between two counterpropagating
edge modes described by chiral Luttinger liquids was calculated
for the case when tunneling takes place along an extended
region.\cite{Aranzana}

In this work we evaluate the backscattering current off {\it
several time-dependent extended impurities} in an otherwise
perfect Luttinger liquid, in the presence of repulsive
electron-electron interactions. We start in Section II with a
single static extended impurity in the noninteracting limit ($g =
1$), which can be solved using the B\"uttiker-Landauer
formalism,\cite{BLandauer} and then we generalize the results for
the case of an arbitrary static potential. In Section III we
include interactions among the electrons and study the problem
using the Keldysh formalism. Within first order perturbation
theory, we first derive a general expression for the
backscattering current in the presence of several time-dependent
extended impurities and then we concentrate on a typical example,
namely, the case of one impurity, to illustrate the main results.
Finally, we compare our findings obtained within perturbation
theory in the limit of noninteracting electrons ($g = 1$) with the
exact results derived in the framework of the B\"uttiker-Landauer
formalism and show a perfect agreement between them. Our
conclusions are presented in Section IV.

\section{Non-interacting electrons ($g=1$) in an extended backscattering potential}
\subsection{The Model}
We start by studying the simplest case, namely, noninteracting
electrons in 1D in the presence of a backscattering potential. The
Hamiltonian of the model reads $H=H_{0}+H_{imp}$, where
\begin{eqnarray}\label{Hamiltonian}
H_{0} &=& -i\hbar v_{F}\int dx (\psi_{R}^{\dagger}
\partial_{x}\psi_{R}-\psi_{L}^{\dagger}
\partial_{x}\psi_{L}), \\
H_{imp} &=& \int dx (\psi_{R}^{\dagger}
\psi_{L}+\psi_{L}^{\dagger} \psi_{R})W(x,t).\nonumber
\end{eqnarray}
Here $\psi_{R/L}$ are  fermionic fields corresponding to the
right/left moving electrons and $W(x,t)$ stands for the
backscattering potential. We also further set the Planck constant
$\hbar=1$ and the Fermi velocity $v_{F}=1$. The equations of
motion then read
\begin{eqnarray}
 (\partial_{t}+\partial_{x})\psi_{R}+i W(x,t)\psi_{L}&=& 0,\\
 (\partial_{t}-\partial_{x})\psi_{L}+i W(x,t)\psi_{R}&=& 0. \nonumber
 \label{EOM}
\end{eqnarray}
 For a static backscattering potential, $W(x,t)=W(x)$, the
 solutions of the above equations of motion have a stationary form,
 $\mathbf{\psi}_{R/L}(x,t)=\mathbf{\psi}_{R/L}(x)e^{-i \omega t}$.
 The fermionic fields $\psi_{R/L}$ then obey the following equation
 of motion
\begin{equation}
\left( \begin{array}{cc} \omega+i\partial_{x} & -W(x)\\
 -W(x) & \omega-i\partial_{x}\\
\end{array}
 \right)\mathbf{\psi}(x)=0,
 \label{SEOM}
\end{equation}
where we introduced a ``spinor'' notation
\begin{eqnarray}
\mathbf{\psi}\equiv \left( \begin{array}{c} \psi_{R} \\
\psi_{L}\\
\end{array}
\right).
\end{eqnarray}
In the following we concentrate on some particular examples for
which an analytical solution can be obtained.
\subsection{Single static rectangular-like barrier}

Let us begin by considering a static rectangular-like potential
barrier of width $L$ and height $W$,
\begin{equation}\label{rectpotential}
W(x)=W[\Theta(x+L/2)-\Theta(x-L/2)],
\end{equation}
 where $\Theta(x)$ is the Heavyside theta-function. The solution of the stationary
equation of motion\ (\ref{SEOM}) may be sought in the form
\begin{eqnarray}
 \label{Eansatz}
 \psi_{R}(x)&=&\psi_{R}^{-}e^{- i k x}+\psi_{R}^{0}e^{i k x }, \\
 \psi_{L}(x)&=&\psi_{L}^{0}e^{- i k x}+\psi_{L}^{-}e^{i k x }.\nonumber
\end{eqnarray}
Substitution of the ansatz\ (\ref{Eansatz}) into the equations of
motion (\ref{SEOM}) yields
\begin{equation}
\psi_{L}^{-}=\frac{W(x)}{\omega+k}\psi_{R}^{0}, \qquad
\psi_{R}^{-}=\frac{W(x)}{\omega+k}\psi_{L}^{0},
\end{equation}
with $\omega^{2}=[W(x)]^{2}+k^{2}$. We now consider the region
$[-L/2,L/2]$ and require both functions $\psi_{L}(x)$ and
$\psi_{R}(x)$ to be continuous at the points $x=\pm L/2$. For the
point $x=-L/2$ we have
\begin{eqnarray}
 \psi_{R}(-L/2)&=&\psi_{R}^{0}e^{-i k L/2 } +
\frac{W}{\omega+k} \psi_{L}^{0}e^{i k L/2 }, \\
 \psi_{L}(-L/2)&=& \frac{W}{\omega+k} \psi_{R}^{0}e^{-i k L/2 }
+\psi_{L}^{0}e^{i k L/2 },\nonumber
 \end{eqnarray}
where $k=\sqrt{\omega^{2}-W^{2}}$, and for the point $x=L/2$ the
expressions are similar. Introducing the matrices
\begin{equation}
B= \left( \begin{array}{cc} 1 & \frac{W}{\omega+k}\\
\frac{W}{\omega+k} & 1
\end{array}
 \right)
\end{equation}
and
\begin{equation}
P= \left( \begin{array}{cc} e^{i k L/2} & 0\\
0 & e^{-i k L/2}
\end{array}
 \right),
\end{equation}
the continuity conditions can be written in a simple form
\begin{equation}
\mathbf{\psi}(-L/2)=BP^{\dagger}\mathbf{\psi}^{0}, \qquad
\mathbf{\psi}(L/2)= BP\mathbf{\psi}^{0}.
\end{equation}
The above equations then yield
 \begin{equation}\label{scattering}
\mathbf{\psi}(L/2)=M\mathbf{\psi}(-L/2), \qquad M=BP^2B^{-1},
\end{equation}
where $M$ is the transmission matrix, which gives the
transformation from the states on the left of the barrier,
$\psi_{R/L} (-L/2)$, to the states on the right of the barrier
$\psi_{R/L} (L/2)$,
\begin{equation}
M= \left( \begin{array}{cc} \cos(k L)+i \frac{\omega}{k} \sin(k L)
&
 -i \frac{W}{k} \sin(k L)\\
i \frac{W}{k} \sin(k L) & \cos(k L)-i \frac{\omega}{k} \sin(k L)
\end{array}
 \right).  \label{Mmatrix}
\end{equation}
Let us observe that
$M^\dagger\mathbf{\sigma}_{3}M=\mathbf{\sigma}_{3}$, and therefore
$\bar{\mathbf{\psi}}\mathbf{\psi}$ is a constant, where
$\bar{\mathbf{\psi}}=\mathbf{\psi}^{\dagger}\mathbf{\sigma}_{3}$
is the Dirac conjugate. This is a consequence of charge
conservation and holds for any transmission matrix, since an
arbitrary backscattering potential can always be represented as a
set of (infinitely narrow) rectangular potentials with different
amplitudes. Therefore, the transmission matrix $M\in SU(1,1)$ can
be always written in the form
\begin{equation}
M= \left( \begin{array}{cc} s & r^{\ast}\\
r & s^{\ast}
\end{array}
 \right).  \label{Melements}
\end{equation}
A more elegant derivation of Eq.\ (\ref{Mmatrix}) using results of
the Lie group theory is presented in the Appendix. The
transmission matrix given by Eq.\ (\ref{Melements}) is related to
the scattering matrix, $S$, which gives the transformation from
the incoming modes $\psi_{R}(-L/2),\psi_{L}(L/2)$ to the outgoing
modes $\psi_{R}(L/2),\psi_{L}(-L/2)$. In terms of the transmission
matrix elements, the latter reads
\begin{equation}
S= \frac{s}{|s|}\left( \begin{array}{cc} 1/|s| & r^{\ast}/|s|\\
-r/|s| & 1/|s|
\end{array}
 \right).  \label{Selements}
\end{equation}

Let us now calculate the transmission coefficient for the right
moving particles using Eq.\ (\ref{scattering}). In that case, we
have
\begin{equation}
\left( \begin{array}{c} \psi_{R}(L/2) \\
0\\
\end{array}
\right)=M\left( \begin{array}{c} \psi_{R}(-L/2) \\
\psi_L(-L/2)\\
\end{array}
\right),
\end{equation}
with the transmission matrix $M$ given by Eq.\ (\ref{Mmatrix}).
The transmission coefficient may be then promptly found from the
previous equation
\begin{equation}
T\equiv\left|\frac{\psi_R(L/2)}{\psi_R(-L/2)}\right|^2=\frac{1}{|s|^2}=
\frac{1}{1+|r|^2},
\end{equation}
Using the explicit form of the transmission matrix, given by Eq.\
(\ref{Mmatrix}), we obtain for $\omega^{2}\geq W^{2}$
\begin{equation} T(\omega)=\left[{1+
\frac{W^{2}}{\omega^{2}-W^{2}} \sin^{2}(\sqrt{\omega^{2}-W^{2}}
L)}\right]^{-1}
 \label{Transm}
\end{equation}
and for $\omega^{2}< W^{2}$
\begin{equation} T(\omega)=\left[{1+
\frac{W^{2}}{W^{2}-\omega^{2}} \sinh^{2}(\sqrt{W^{2}-\omega^{2}}
L)}\right]^{-1}.
 \label{Transm2}
\end{equation}
An analogous calculation shows that the transmission coefficient
for the left moving particles is the same as for the right moving
ones. We can now find the backscattering current using the
Landauer formula\cite{BLandauer} and choosing the left and right
reservoirs to be symmetric. We use the subscript ``$st$'' to
denote the backscattering current off a {\it static} impurity to
distinguish it from the backscattering current originating from
{\it time-dependent} {\it barriers}, which is considered in the
next section. The current then reads
\begin{equation}
I_{st}=e\int_{-\infty}^{\infty} \frac{d\omega}{2\pi}
[n^{R}(\omega)-n^{L}(\omega)][T(\omega)-1], \label{BLform}
\end{equation}
where $n^{R/L}(\omega)$ are the occupation numbers of the
right/left movers, given by the Fermi distribution function
\begin{equation}
n^{R/L}(\omega)=\frac{1}{e^{\beta(\omega-\omega_{F}\mp
\omega_{0}/2)}+1}.
\end{equation}
Here, $\omega_{0}=eV$ is the Josephson frequency related to the
applied bias voltage $V$ and $\omega_{F}$ is the Fermi energy
(recall that $\hbar$=1). We have to distinguish the two cases
given by Eqs.\ (\ref{Transm}) and (\ref{Transm2}), since if
$\omega^{2}< W^{2}$ the momentum $k$ becomes imaginary and
transport occurs then only via tunnelling. In other words, the
presence of backscattering opens a gap in the energy spectrum of
the fermions from $\omega_F-W$ to $\omega_F+W$. Substituting Eq.\
(\ref{Transm}) into Eq.\ (\ref{BLform}) and keeping only the
lowest order terms in $W$, we find the ``static'' backscattering
current
\begin{equation}
I_{st}=-\frac{eW^{2}}{2\pi}\int_{-\infty}^{\infty}
d\omega\frac{\sin^{2}(\omega
L)}{\omega^{2}}[n^{R}(\omega)-n^{L}(\omega)]. \label{Tbscurr}
\end{equation}
At zero temperature ($T=0$) and $\omega_{F}=0$ the above
expression simplifies to
\begin{equation}
I_{st}=-\frac{2eW^{2}L}{\pi}\int_{0}^{L\omega_{0}} d y
\frac{\sin^{2}(y/2)}{y^{2}}. \label{bscurr}
\end{equation}
\begin{figure}[tbh]
\begin{center}
\includegraphics[width=6cm,angle=-90]{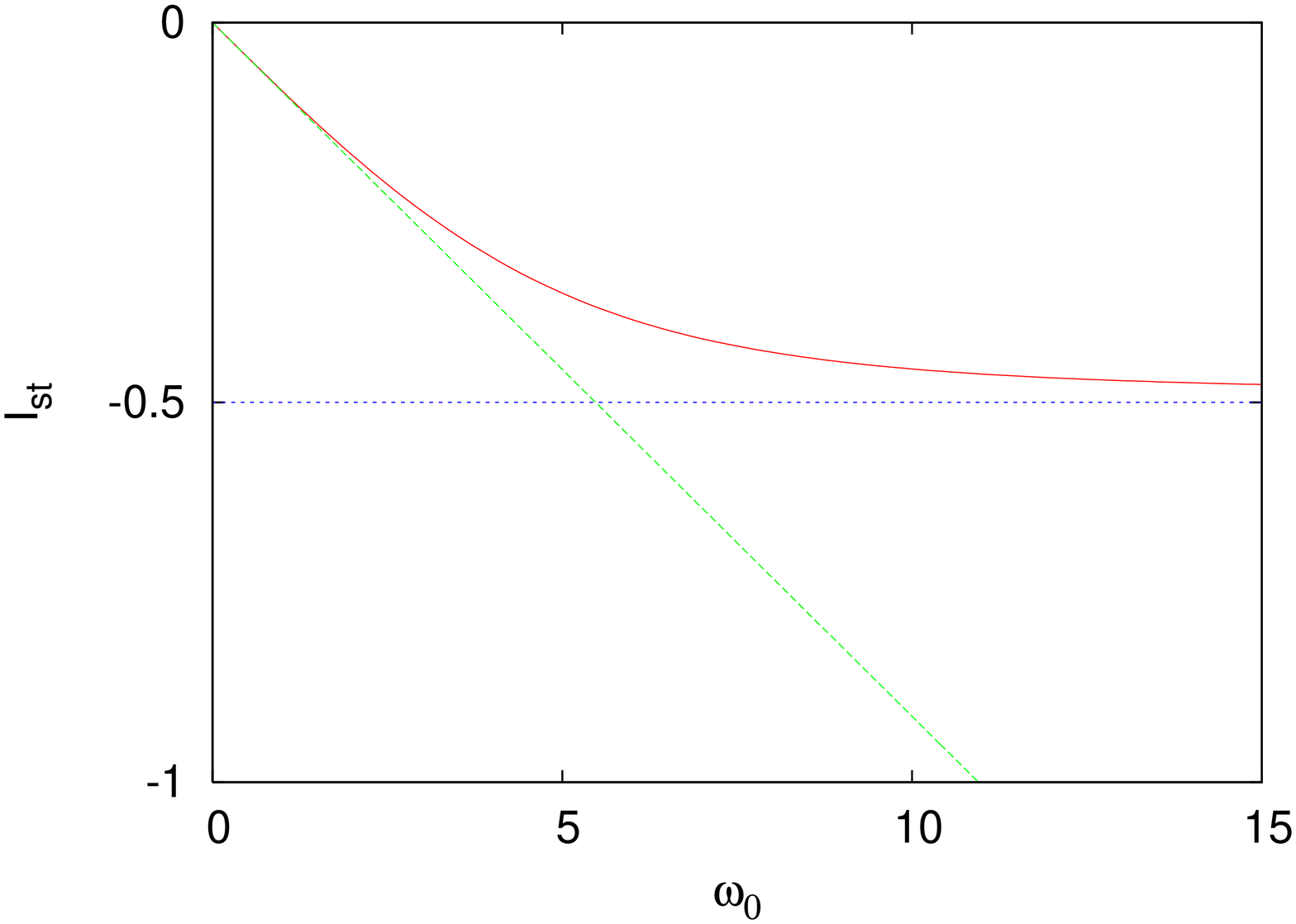}
\includegraphics[width=6cm,angle=-90]{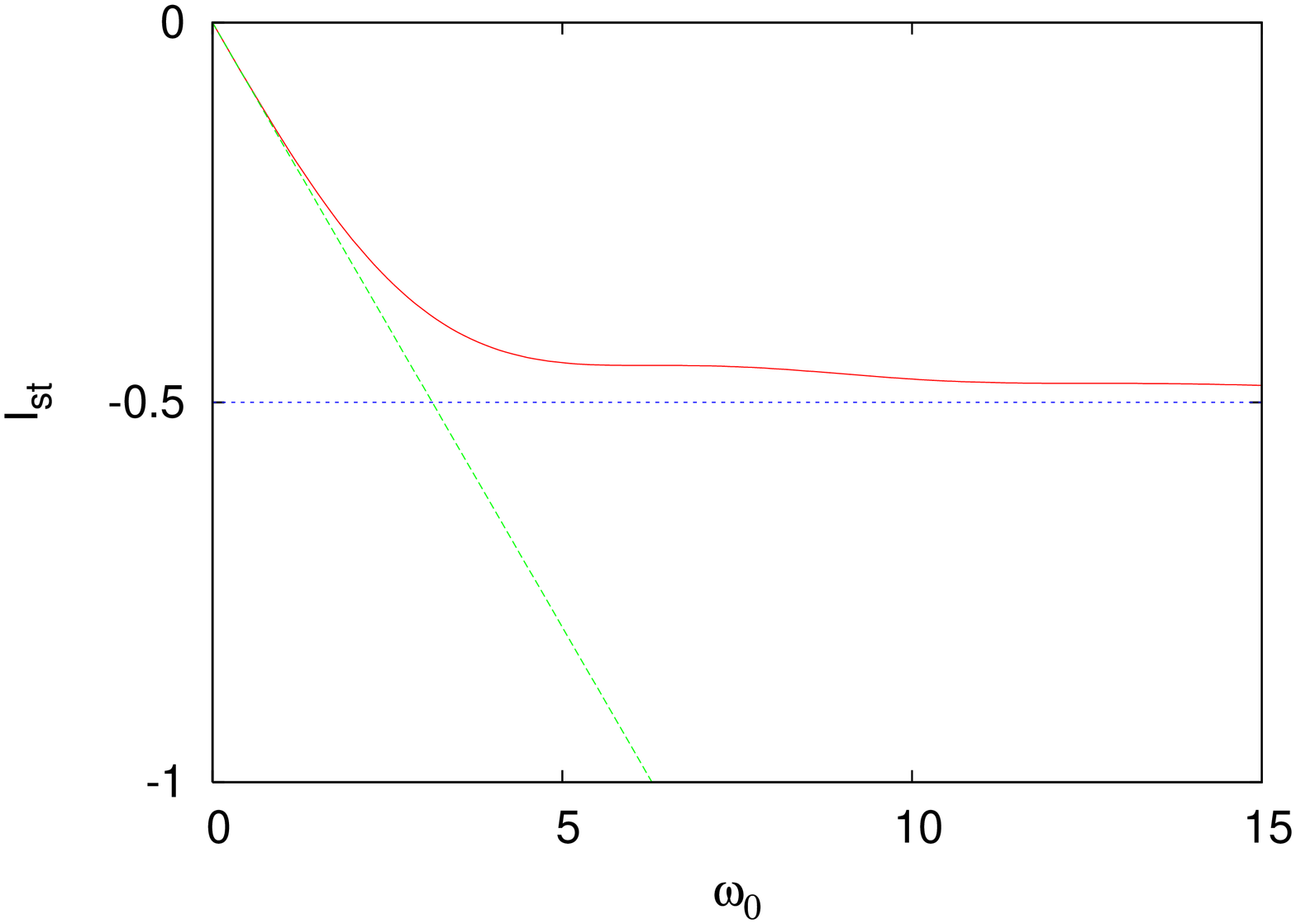}
\end{center}
\caption{\label{fig1}(color online) $I_{st}$ versus $\omega_0$ for
noninteracting electrons ($g=1$) in the presence of a point-like
(dashed line) or an extended impurity (solid line). The asymptotic
behavior for the latter,
$I_{st}(\omega\rightarrow\infty)\rightarrow-1/2$, is represented
by dotted lines. The current is plotted in units of $eW^{2}L$ and
the frequency is in units of $1/L$. Upper panel: $T=1$; lower
panel: $T=0$. }
\end{figure}

The backscattering currents at a finite and zero temperature,
given by Eqs.\ (\ref{Tbscurr}) and (\ref{bscurr}), are plotted,
respectively, in the upper and the lower panel of Fig.\ 1. In both
cases the current $I_{st}$ is represented in units of $eW^2L$ and
the Josephson frequency is in units of $1/L$. Inspection of Fig.\
1 shows that in the region $L\omega_{0}<2$, the current $I_{st}$
for scattering off an extended impurity (solid line) is almost
indistinguishable from the backscattering current of a point-like
impurity represented by a barrier  ${\tilde W}(x)=W L\delta(x)$
(dashed line). However, in the region $L\omega_{0}>2$ the
difference becomes drastic, since the backscattering current off a
point-like impurity continues to decrease linearly, whereas for
the extended impurity, the current saturates at the value $I_{bs}=
-e W^{2}L/2$.

\subsection{Arbitrary static potential}

 Let us now consider noninteracting electrons backscattered
off a static potential $W(x)$ localized in the interval $0\leq
x\leq y$ with the corresponding transmission matrix $M(y)$. In
order to find the dependence of the transmission matrix on the
scattering potential, let us add to the existing barrier an
infinitesimally thin part with the height $W(y+dy)$.
In this case the transmission matrix at the point $y+dy$ is
$M(y+dy)=M(y)+dM$, where $dM=i{\cal T}M dy$ (see Appendix). After
performing a matrix multiplication, we find the following
differential equations
\begin{eqnarray}
 d s=i[\omega s - W(y) r ]dy, \\ \nonumber
 d r=i[W(y) s - \omega r]dy.
\end{eqnarray}
If the potential is weak, $W(y) \ll \omega$, then to the lowest
order in perturbation expansion $r={\cal O} (W/\omega)$. After
neglecting terms ${\cal O} (r^{3})$, we obtain
\begin{equation}
d r(y)=i[W(y)e^{i \omega y}- \omega r(y)]dy.
\end{equation}
This is a linear inhomogeneous differential equation of the first
order with solution of the form
\begin{equation}
r(y)=ie^{-i \omega y}\int_{0}^{y}dxW(x)e^{i 2\omega x}.
\end{equation}
Then, it follows straightforwardly that
\begin{equation}
|r|^2=\int dx_{1}\int dx_{2}W(x_{1})W(x_{2})\cos[ 2\omega
(x_{1}-x_{2})].
\end{equation}
On the other hand, since $|r|^2\ll 1$, the transmission
coefficient becomes
\begin{equation}
T(\omega)-1=\frac{1}{1+|r|^2} - 1 =-|r|^2,
\end{equation}
and the  backscattering current at finite temperatures then reads
\begin{equation}
I_{st}=-e\int_{-\infty}^{\infty} \frac{d\omega}{2\pi}
|r|^2[n^{R}(\omega)-n^{L}(\omega)].
\end{equation}
Consequently, we find that
\begin{eqnarray}
&& \frac{\delta^{2}}{\delta W(x_{1})\delta
W(x_{2})}I_{st}{\Large{\vert}}_{W=0}=\\ \nonumber &&
-e\int_{-\infty}^{\infty} \frac{d\omega}{\pi}\cos[2\omega
(x_{1}-x_{2})][n^{R}(\omega)-n^{L}(\omega)].
\end{eqnarray} Since
\begin{equation}\nonumber
\int_{-\infty}^{\infty}d y\cos(y) [n^{R}(y/2l)-n^{L}(y/2l)]=4 \pi
T l\frac{\sin(\omega_0 l)\cos(2\omega_{F}l)}{\sinh(2 \pi T l)},
\end{equation}
we eventually obtain that
\begin{eqnarray}
 \label{twopointnonint} && \frac{\delta^{2}I_{st}}{\delta
W(x_{1})\delta W(x_{2})}{\Large{\vert}}_{W=0}=\\ \nonumber &&
-2eT\frac{\cos[2\omega_{F}(x_{1}-x_{2})]\sin[\omega_{0}(x_{1}-x_{2})]}{\sinh[2T\pi
(x_{1}-x_{2})]}.
\end{eqnarray}
As we will see later, Eq.\ (\ref{twopointnonint}) will allow us to
connect the results obtained within perturbation theory in the
bosonization formalism with the ones derived in the framework of
the B\"uttiker-Landauer approach, for an arbitrary (but weak)
extended impurity barrier, in the limit of noninteracting
electrons ($g=1$).

\section{Interacting electrons}
\subsection{Several time-dependent extended
impurities} In this section we consider a one-dimensional quantum
wire with spinless interacting electrons backscattered by a {\it
time-dependent} potential $W(x,t)$. The Hamiltonian of the model,
including the electron-electron interaction, reads
\cite{FeldmanGefen,Chamon2,Rao}
\begin{eqnarray}\nonumber
H &=&\int dx \left\{-i(\psi_{R}^{\dagger}
\partial_{x}\psi_{R}-\psi_{L}^{\dagger}\partial_{x}\psi_{L})
+ U(\psi_{R}^{\dagger}\psi_{R}+\psi_{L}^{\dagger} \psi_{L})^{2}
\right. \\  &+& \left. W(x,t)\left[\psi_{L}^{\dagger}
\psi_{R}e^{i\omega_{0}t+i2k_{F}x}+H.c.\right]\right\},
\label{Hmodel}
\end{eqnarray}
where $U$ is the interaction strength. In the next step, we
consider a discretized backscattering potential $W(x,t)$ composed
of several point-like time-dependent impurities located at
positions $x_p$, $p=1,...,N$, such that
 \begin{equation}
 \lim_{\Delta
x_p\rightarrow 0+}\sum_{p=1}^{N}W(x_p,t)\Delta x_{p} = \int W(x,t)
dx.
\end{equation}

We now use the bosonization technique, which relates the fermionic
fields, $\psi_{R/L}$, with the corresponding bosonic fields,
$\Phi_{R/L}$, through $\psi_{R/L}=\alpha \eta e^{\pm
i\sqrt{g}\Phi_{R/L}}$. Here, $\alpha$ is a normalization factor,
$\eta$ is a fermionic operator, and $g=\sqrt{\pi /(\pi+2U)}$
(recall $v_{F}=1$) is the coupling constant in the bosonic theory.
The action corresponding to the Hamiltonian\ (\ref{Hmodel}) then
reads \cite{FeldmanGefen,Chamon,Chamon2,Rao,us,Egger}
\begin{eqnarray}
S &=&\int dt dx \left\{\frac{1}{8\pi}\left[(\partial_{t}\Phi)^{2}-
(\partial_{x}\Phi)^{2}\right] \right. \\ \nonumber &-&
\left.\sum_{p}\delta(x-x_{p})V_{p}(t)\left(e^{i\sqrt{g}
\Phi(t,x_{p})} e^{i\omega_{0}t}+H.c.\right)\right\},
\end{eqnarray}
where $V_p(t)\equiv W(x_p,t)e^{i2k_{F}x_p}\Delta x_p/2\pi$ and
$\Phi\equiv\Phi_R+\Phi_L$. Using the approach developed in Refs.\
[\onlinecite{us}] and [\onlinecite{Chamon2}], we find that, to the
lowest order in $V$, the backscattering current reads
 \begin{equation}
 I_{bs}(\omega_{0},t)=C \sum_{k,j} \int_{|x_{k}-x_{j}|}^{\infty}
 \frac{d\tau\sin(\omega_{0}\tau)  V^{\ast}_{k}(t-\tau)
 V_{j}(t)}{|\tau^{2}-(x_{k}-x_{j})^{2}|^{g}}+H.c.,
\label{Ibs}
\end{equation}
where $C = -2 e\sin (\pi g)$. The most general case should involve
a spatially varying impurity potential, with an arbitrary shape.
In order to account for this more general situation, we take the
continuum limit in Eq.\ (\ref{Ibs}) by setting the number of
impurities $N \rightarrow \infty$ and the distance between the
impurities $\Delta x_p\rightarrow0$. Eq.\ (\ref{Ibs}) then
acquires the form
\begin{eqnarray} \nonumber
 I_{bs}(\omega_{0},t)=-e\frac{\sin (\pi g)}{\pi^{2}} \int\int dx_{1}
dx_{2}\int_{|x_{1}-x_{2}|}^{\infty}d\tau \\
\frac{\sin(\omega_{0}\tau)
\cos[2k_{F}(x_{1}-x_{2})]W(x_{1},t-\tau)W(x_{2},t)}
 {|\tau^{2}-(x_{1}-x_{2})^{2}|^{g}}.
 \label{extimp}
 \end{eqnarray}
Eq.\ (\ref{extimp}) is a very important result of our paper
because it allows us to evaluate the backscattering current for
any set of time-dependent extended impurities, described by any
arbitrary function $W(x,t)$.

We can progress further in the analytical calculation of some
specific examples by introducing a simplifying assumption, namely,
that the impurity potential is a periodic function of time,
$W(x,t)=W(x)\cos(\Omega t)$. We have shown in Ref.\
\onlinecite{us} that the dc component of the backscattering
current generated by a set of dynamical point-like impurities
oscillating with a frequency $\Omega$ can be expressed, in first
order of perturbation theory, in terms of the dc current for a set
of quasi-static impurities ($\omega_{0} \gg \Omega >
1/\tau_{meas}$, with $\tau_{meas}$ being the measuring or the
relaxation time),
\begin{equation}
I_{dc}=\frac{1}{2}[I_{qst}(\omega_{0}+\Omega)+I_{qst}(\omega_{0}-\Omega)],
\label{Idc}
\end{equation}
or static impurities ($1/\tau_{meas}\gg\Omega$)
\begin{equation}
I_{dc}=\frac{1}{4}[I_{st}(\omega_{0}+\Omega)+I_{st}(\omega_{0}-\Omega)],
\label{Idc2}
\end{equation}
with
\begin{eqnarray} \nonumber
I_{st}(\omega_0)=-e\int\int dx_{1}
dx_{2}W(x_{1})W(x_{2})\\
\cos[2k_{F}(x_{1}-x_{2})]H_{g}(\omega_{0},|x_{1}-x_{2}|,T = 0).
\label{Idcstatic}
\end{eqnarray}
The function $H_{g}$ is defined as
\begin{equation}
H_{g}(\omega,x,T=0) \equiv \frac{\sin (\pi g)}{\pi^{2}}
\int_{|x|}^{\infty}d\tau \frac{\sin(\omega
\tau)}{|\tau^{2}-x^{2}|^{g}} \label{Ffunction}
\end{equation}
and after explicit evaluation it reads
\begin{equation}
H_{g}(\omega,x,T=0)={\rm {\rm
sgn}}(\omega)\frac{J_{g-1/2}(x\omega)}{2^{1/2+g}\sqrt{\pi}
\Gamma(g)}\left|\frac{\omega}{x}\right|^{g-1/2},
\end{equation}
where $J_{\alpha}(x)$ are Bessel functions of the first kind and
$\Gamma(g)$ is the gamma function.

This result may be readily generalized for the situation where a
phase shift of the form $W(x,t)=W(x)\cos(\Omega t+2Kx)$ is present
along an extended impurity. Then we can again write the dc
component of the current as
\begin{equation}
I_{dc}=\frac{1}{4}[I_{st,k_{F}+K}(\omega_{0}+\Omega)+I_{st,k_{F}-K}(\omega_{0}-\Omega)],
\end{equation}
where the channels $\omega_{0}+\Omega$ and $\omega_{0}-\Omega$
have an ``effective'' Fermi level at the momentum $k_{F}+K$ and
$k_{F}-K$, respectively.
This means that it is possible to decrease the ``effective'' Fermi
momentum for the $\omega_{0}-\Omega$ channel. For simplicity we
assume $K=0$ in the following.

We discussed until now the $T=0$ case. When the temperature is
nonzero, the Green's function is modified by a conformal
transformation \cite{Shankar}
\begin{equation}
|\tau^{2}-x^{2}|^{g}\rightarrow \frac{|\sinh[\pi
T(\tau-x)]\sinh[\pi T(\tau+x)]|^{g}}{(\pi T)^{2g}}.
\end{equation}
 Consequently, the function $H_{g}$ acquires the form
\begin{equation}
H_{g}(\omega,x,T) \equiv \int_{|x|}^{\infty}\frac{d\tau \sin (\pi
g)\sin(\omega \tau)T^{2g}\pi^{2g-2}}{|\sinh[\pi
T(\tau-x)]\sinh[\pi T(\tau+x)]|^{g}}. \label{FfunctionwithT}
\end{equation}

It is worth noting here that the function $I_{st}(\omega_0)$,
defined by Eq.\ (\ref{Idcstatic}), can be represented as a product
of a dimensional factor(which actually becomes dimensional only
after including the cutoff of the model) and a function of
dimensionless variables. The dimensionless part remains invariant
with scaling
\begin{eqnarray} \nonumber
\label{scaling} \omega_0\rightarrow \omega'_0 = \omega_0 \zeta,
\quad x\rightarrow
x'=x/\zeta,\\
\quad W\rightarrow W' = W \zeta, \quad T\rightarrow T' = T \zeta,
\end{eqnarray}
whereas the currents $I_{st}$ and $I_{dc}$ scale as the
dimensional factor
\begin{eqnarray} \nonumber
I_{st}&\rightarrow& I'_{st}\\
I'_{st}[W'(x')](\omega_0',T')&=&\zeta^{2g-1}
I_{st}[W(x)](\omega_0,T). \label{scaling2}
\end{eqnarray}
These equations set the relation between currents corresponding to
different potentials, which have the same shape but a different
length parameter. For example, all rectangular-like potentials
have the same backscattering current dependence up to the scaling
(\ref{scaling}). This leads us to the conclusion that it is enough
to consider some potential with a fixed length parameter to obtain
the backscattering current dependence for all potentials of the
same form (in lowest order of perturbation theory).
\begin{figure}[htb]
\begin{center}
\includegraphics[width=6cm,angle=-90]{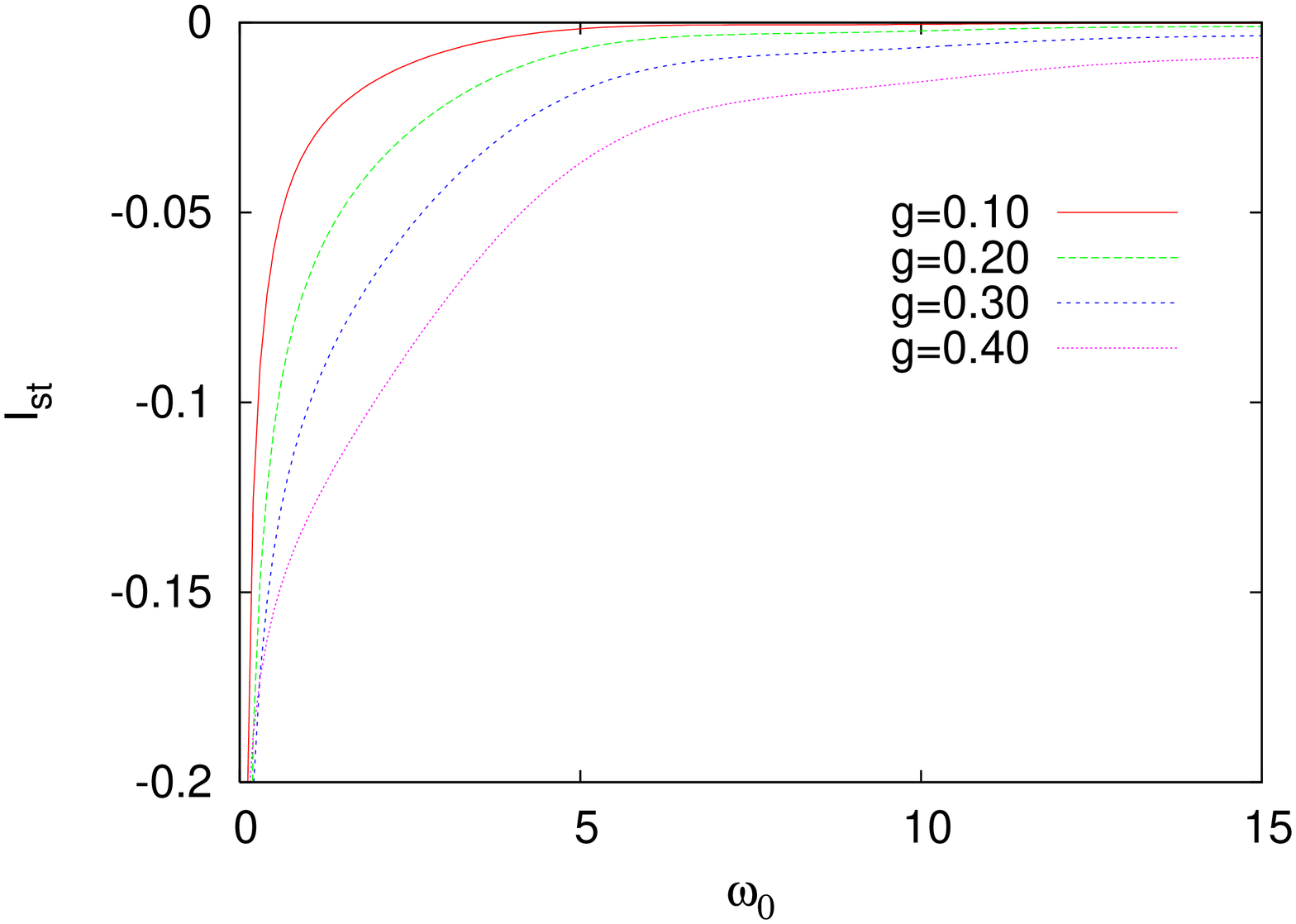}
\includegraphics[width=6cm,angle=-90]{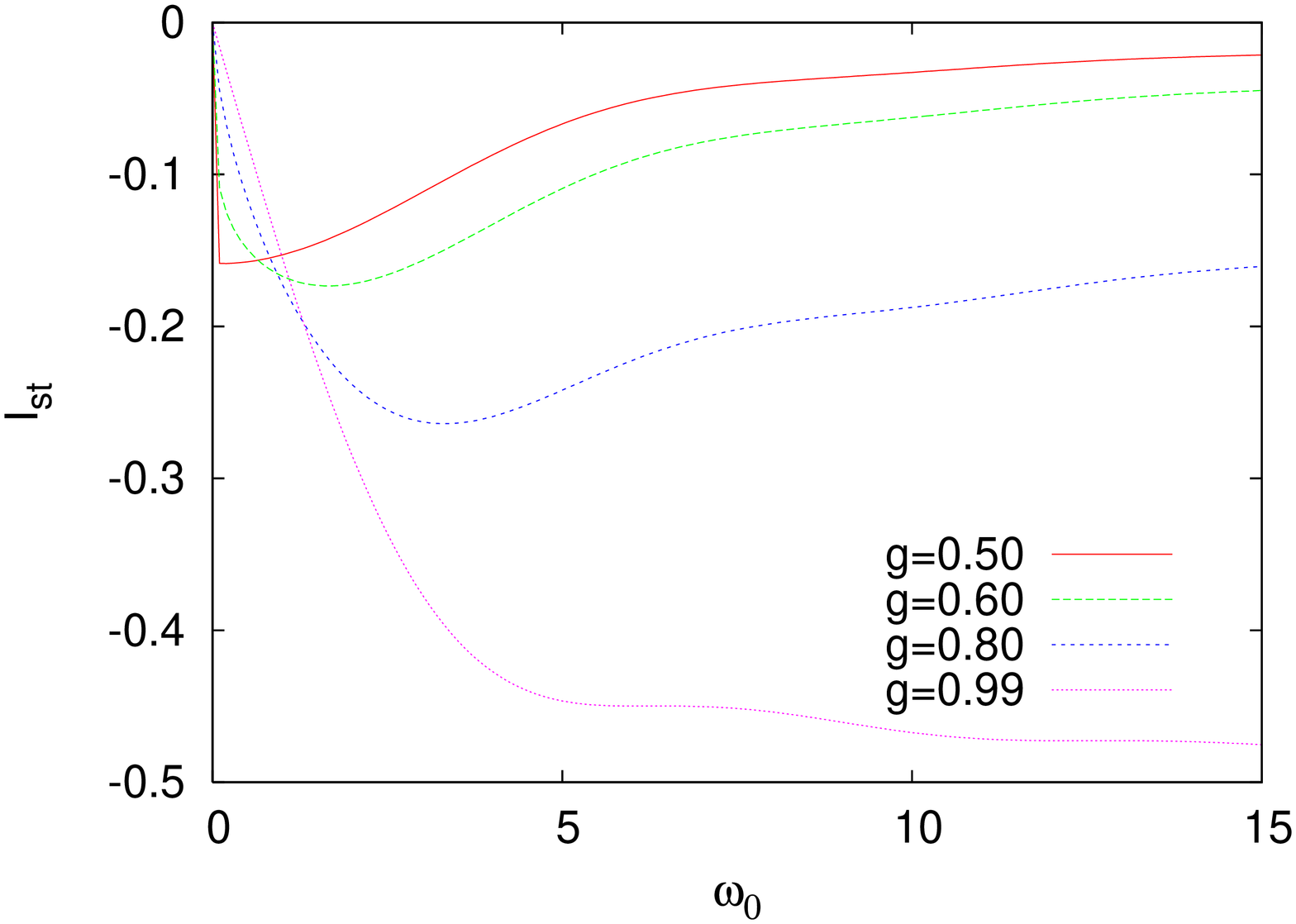}
\end{center}
\caption{\label{fig2}(color online) Dimensionless "static" current
$I_{st}$ versus $\omega_0$ for $k_{F}=0$ and $W=L=1$. Upper panel:
$g<1/2$. Lower panel: $g\geq1/2$. }
\end{figure}

In the case of a potential $W(x)$ with a  rectangular shape, the
above integral Eq.\ (\ref{Idcstatic}) simplifies to
\begin{eqnarray} \nonumber
 I_{st}(\omega_0)&=&-eW^{2}\int_{-L/2}^{L/2}\int_{-L/2}^{L/2} dx_{1}
dx_{2}\cos[2k_{F}(x_{1}-x_{2})]\\ \nonumber
&&H_{g}(\omega_{0},|x_{1}-x_{2}|,T).\end{eqnarray} By changing the
variables $x=x_{1}-x_{2}$ and $y=x_{1}+x_{2}$, one integration can
be performed, finally yielding
\begin{equation}\label{Irect}
I_{st}(\omega_0)=-2eW^{2}\int_{0}^{L} dx (L-x)\cos(2k_{F}
x)H_{g}(\omega_{0},x,T).
\end{equation}

In Fig.\ 2, the dimensionless current in the presence of a static
impurity, given by Eq.\ (\ref{Irect}), is plotted for $k_{F}=0$,
$W=L=1$, since, as it was argued before, we can fix the length of
the barrier without loss of generality. The upper panel shows that
for $g<1/2$, $I_{st}$ always increases, analogous to the case of a
point-like impurity. On the other hand, for $g>1/2$, a
non-monotonic behavior of the current appears. As we will discuss
later, this non-monotonic behavior may lead to an increase of the
conductance, beyond the value $G_0$.

Let us now discuss the dependence of the current on the value of
$k_F$. To demonstrate the influence of the Fermi momentum, $k_F$,
on the backscattering current off a static impurity, we plot
$I_{st}$ versus the Josephson frequency for several values of the
Fermi momentum in Fig.\ 3.
\begin{figure}[tbh]
\begin{center}
\includegraphics[width=8cm,angle=0]{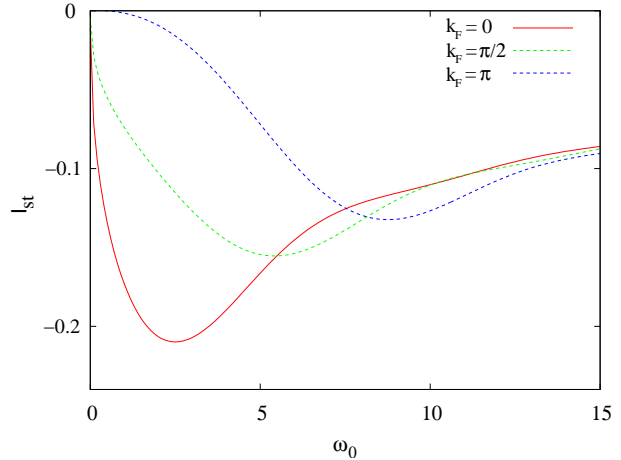}
\end{center}
\caption{\label{fig3}(color online) Dimensionless $I_{st}$ versus
$\omega_0$ for $g=0.7$ and $L=1$}
\end{figure}
It can be promptly observed that the qualitative features are the
same for zero and finite $k_{F}$, the only effect of the latter is
to shift the minimum of the curve to higher values of
$\omega_{0}$.

For future purposes it is worth noting that for the rectangular
potential with $k_{F}=0$ the following relation holds
\begin{equation}\label{partialst}
\frac{\partial^{2}}{\partial
L^{2}}I_{st}=-2eW^{2}H_{g}(\omega_{0},L,T).
\end{equation}
Eq.\ (\ref{partialst}) is a consequence of a more general
relation, valid for an arbitrary static potential \cite{Liguori}
\begin{eqnarray} \nonumber
&&\frac{\delta^{2}}{\delta W(x_{1})\delta
W(x_{2})}I_{st}\vert_{W=0}= \\ \nonumber
&&-2e\cos[2k_{F}(x_{1}-x_{2})]H_{g}(\omega_{0},|x_{1}-x_{2}|,T).
\end{eqnarray}

\subsection{Noninteracting electrons $g=1$ within perturbation theory}

Now we can compare this result with the one obtained in Sec. II B
using the B\" uttiker-Landauer formalism by taking the limit
$g\rightarrow1$ in Eq.\ (\ref{Ffunction}). Defining
$h(\tau)\equiv\sin(\omega\tau)$, $f(\tau)\equiv|\tau^2-x^2|$, and
$g=1-\epsilon$, we may use the formula
\begin{equation}
\lim_{\epsilon\rightarrow 0+}\int_{x}^{y}d \tau
h(\tau)[f(\tau)]^{\epsilon-1}\epsilon=\frac{h(x)}{f'(x)}, \quad
y>x,
\end{equation}
which holds for a function $f$ with the properties $f(x)= 0$ and
$f'(x)>0$, yielding
\begin{equation}
H_{1}(\omega_{0},x,T)=T\frac{\sin(\omega_{0}x)}{\sinh(2T\pi x)}.
\label{H1}
\end{equation}

In the case of a rectangular potential and $k_{F} = 0$, we may
substitute Eq.\ (\ref{H1}) into Eq.\ (\ref{Irect}) to obtain
\begin{equation}
I_{st}(\omega_0)=-2eW^{2}T\int_{0}^{L} dx
(L-x)\frac{\sin(\omega_{0}x)}{\sinh(2T\pi x)}, \label{TIst}
\end{equation}
which, in the limit $T\rightarrow0$, becomes
\begin{equation} I_{st}(\omega_0)=-\frac{eW^{2}}{\pi}\int_{0}^{L}
dx (L-x)\frac{\sin(\omega_0 x)}{x}.
\end{equation}
After integrating by parts and using that
$\frac{d}{dy}[2\sin^{2}(y/2)]=\sin(y)$, we find
\begin{equation}
I_{st}=-\frac{2eW^{2}L}{\pi}\int_{0}^{L\omega_{0}} d y
\frac{\sin^{2}(y/2)}{y^{2}},
\end{equation}
which exactly coincides with Eq.\ (\ref{bscurr}) found in the
previous section.

To demonstrate the equivalence between the results obtained using
different approaches in the case of nonzero temperature, we note
that the second derivative of the currents given by Eqs.\
(\ref{Tbscurr}) and (\ref{TIst}) actually coincide,
\begin{equation}
\frac{\partial^{2}}{\partial L^{2}}I_{st}=-2e W^{2}T
\frac{\sin(\omega_0 L)}{\sinh(2 \pi T L)}.
\end{equation}
Therefore,  these currents can only differ by a function
$F(L,\omega_0,T)=C_{1}(\omega_0,T)L+C_{2}(\omega_0,T)$, with $C_1$
and $C_2$ being arbitrary functions of their arguments. It follows
from $\frac{\partial}{\partial
L}I_{st}(L=0,\omega_0,T)=I_{st}(L=0,\omega_0,T)=0$, that
$F(L,\omega_0,T)=0$, and the currents given by Eqs.\
(\ref{Tbscurr}) and (\ref{TIst}) are thus equal. The same
arguments hold also for the case of an arbitrary potential and
$k_{F} \neq 0$, where the equations have similar form, only with
 the partial derivative replaced by the functional one,
$\frac{\partial}{\partial L}\rightarrow \frac{\delta}{\delta
W(x)}$.
\begin{figure}[tbh]
\begin{center}
\includegraphics[width=6cm,angle=-90]{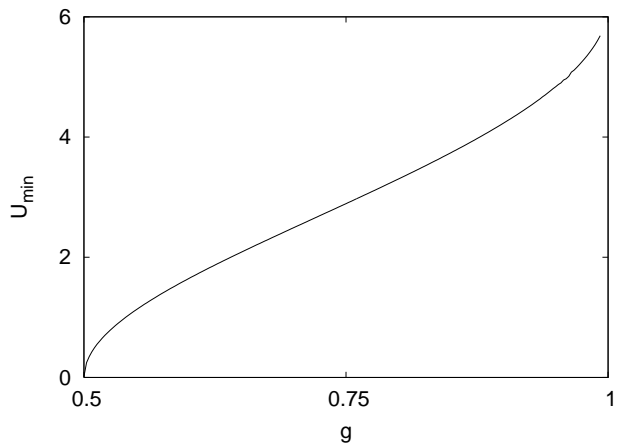}
\end{center}
\caption{\label{fig4}$\omega_{min}L$ versus $g$. }
\end{figure}

\subsection{One oscillating extended impurity}
We can now fix the width of the impurity to be $L=1$ without loss
of generality, since the other cases are obtained from the latter
by the scaling relation (\ref{scaling}). The function
$I_{st}(\omega)$ is odd, $I_{st}(-\omega)=-I_{st}(\omega)$,
because there is no ratchet effect present in the lowest order of
perturbation expansion.\cite{Vinokur} The current for the
dynamical impurity $I_{dc}$, according to Eq.\ (\ref{Idc2}), reads
\begin{equation}
I_{dc}=\frac{1}{4}[I_{st}(\Omega+\omega_0)-I_{st}(\Omega-\omega_0)].
\end{equation}
Since $I_{st}(\omega_0)$ is always negative or zero for
$\omega_0>0$, the function $I_{dc}$ is also negative if both
$\omega_0+\Omega$ and $\omega_0-\Omega$ are positive. To obtain a
positive backscattering current it is therefore necessary to have
$\Omega>\omega_0$, i.e., the frequency of the impurity should be
larger than the external bias frequency. Then for small $\omega_0$
\begin{equation}
I_{dc}=\frac{1}{2}I'_{st}(\Omega)\omega_0 \label{deriv}
\end{equation}
and therefore, the $I_{dc}$ current is positive if
$I'_{st}(\Omega)>0$, at least for small $\omega_0$. Let us begin
by considering the current in the region $1/2<g<1$. For a fixed
value of the interaction parameter $g$ within this region, the
``static'' current $I_{st}(\omega_0)$ continuously decreases from
zero, for $\omega_0=0$, to a minimal value $I_{st}(\omega_{min})$,
and then in the remaining region it increases, see Fig.\
(\ref{fig3}). Let us concentrate in the following on the example
with $k_{F}=0$, which provides a lower bound for $\omega_{min}$.
\begin{figure}[tbb]
\begin{center}
\includegraphics[width=6cm,angle=-90]{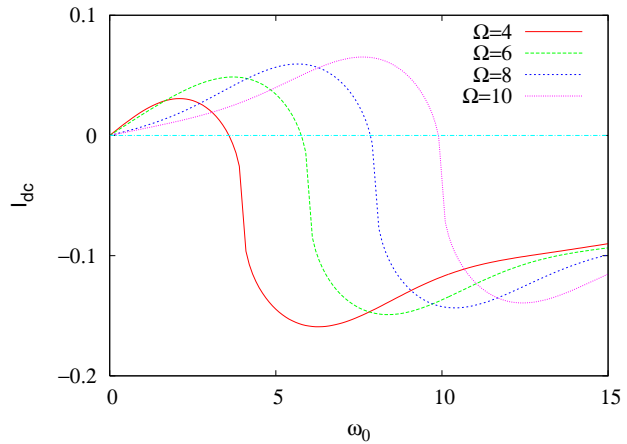}
\end{center}
\caption{\label{fig5}(color online) $I_{dc}$ versus $\omega_0$ for
$g=0.7$. }
\end{figure}
The value of $\omega_{min}L$ then depends only on the interaction
strength $g$, and it changes monotonously from 0 at $g=1/2$ to 5
at $g\approx0.95$, but then diverges to $+\infty$ when
$g\rightarrow 1$, see Fig.\ (\ref{fig4}). Therefore, for typical
values of $g$, $\omega_{min}L$ is of order 1. Note that the
minimum is very sensitive to the impurity length, and never
achievable for the case of a point-like impurity ($L\rightarrow
0$), for which $I_{st}(\omega_{0})$ is a monotonously decreasing
function. Since in our case  the backscattering current changes
sign only when $I'_{st}(\Omega)>0$, we have to choose
$\Omega>\omega_{min}$ in order to obtain an increase of the dc
current, see Eq.\ (\ref{deriv}) and Fig.\ (\ref{fig3}). Now,
considering $I_{dc}$ as a function of the applied voltage $V$ for
some fixed $\Omega>\omega_{min}$, one finds that the dc current is
positive in a region $0<\omega_0 < \omega_{0}^*(\Omega)$, see
Fig.\ (\ref{fig5}). Therefore, the effect of current enhancement
in the region $1/2<g<1$ occurs only if the frequency of the
impurity is larger than some minimal value $\omega_{min}$, which
depends on $g$. For the frequency $\Omega>\omega_{min}$, the
continuous function $I_{dc}$ is then positive within a region $0<
\omega_0 < \omega_{0}^*$.

\section{Conclusions}
In this work we calculated the backscattering current off extended
impurities in a Luttinger liquid. We started our analysis by
considering noninteracting electrons ($g=1$) scattered by a
rectungular-like barrier of width $L$ and height $W$. In this
simplified case the problem can be exactly solved using the
B\"uttiker-Landauer formalism. We evaluate the backscattering
current at finite and zero temperatures for the Fermi momentum
$k_F=0$, and find that for small Josephson frequencies
$L\omega_{0}<2$ the result is almost indistinguishable from the
backscattering current off a point-like impurity represented by a
barrier ${\tilde W}(x)=W L\delta(x)$. In the region
$L\omega_{0}>2$ these currents differ considerably, since in the
case of a point-like impurity the current decreases linearly
whereas for the extended barrier it saturates at the value
$I_{bs}= -e W^{2}L/2$. We also calculated the backscattering
current off a weak backscattering potential with an arbitrary
shape in lowest order of perturbation expansion.

The effect of a time-dependent extended barrier, $W(x,t)$, in the
presence of repulsive electron-electron interactions was then
considered. Using the bosonization technique combined with the
Keldysh formalism, we showed that the backscattering current off
dynamical impurities can be expressed in terms of the
backscattering current off static ones. It actually turns out that
the only effect of the barrier oscillating with a frequency
$\Omega$ is to split the conduction channel $\omega_0$ into two
channels, $\omega_0\pm\Omega$.
We then evaluated the backscattering current off static barriers
$I_{st}(\omega_{0})$ and showed that it is a non-monotonic
function of $\omega_{0}$, which for $1/2<g<1$ has a minimum at
$\omega_{min}$. In addition, we observed that $\omega_{min}$
depends on the interaction strength and the parameter $k_{F}L$,
with $k_{F} = 0$ providing a lower bound for $\omega_{min}$.
Therefore, by choosing $\Omega$ such that $\Omega>\omega_{min}$,
it may occur that the backscattering current changes sign and in a
certain region of applied voltage $0< \omega_0 < \omega_{0}^*$ the
conductance will be greater than the quantum of conductance $G_0$.
The phenomenon of the conductance enhancement may occur in the
presence of oscillating impurities as well, but the main
difference concerns the region of parameters where it appears,
which depends on the type of the barrier. In the case of a single
point-like oscillating barrier it only occurs for strongly
repulsive interactions $0<g<1/2$,\cite{FeldmanGefen} whereas for
two or more point-like barriers it may happen in the entire range
of repulsive interactions $0<g<1$ if the barriers oscillate with
the same frequency $\Omega$.\cite{us} On the other hand, if the
barrier is extended, the phenomenon already appears for {\it a
single} oscillating barrier in the entire region of the repulsive
electron-electron interactions ($0<g<1$). The interval of voltage
where the conductance increases can be optimally chosen by
adjusting the frequency $\Omega$ of the oscillating barrier.

In summary, we investigated the transport properties of a
Luttinger liquid accounting for the most general case of
scattering off extended time-dependent barriers and determined the
values od parameters where the conductance may increase beyond its
quantized value.

\section{Acknowledgements}
The authors have benefitted from fruitful discussions with M. P.
A. Fisher, J. S. Caux, and H. T. C. Stoof.

\appendix

\section{}
The matrix $M$ for a single extended impurity can also be found in
a different approach, using elements of the Lie group theory.
Considering the extended impurity as a set of infinitely narrow
impurities, the matrix $M$ may be represented as an infinite
product of the matrices corresponding to the scattering by the
individual impurities. The transmission matrix corresponding to a
vanishingly narrow potential (Eq.\ (\ref{rectpotential}) with
$L\rightarrow0$) is given by
\begin{equation}
M=I+i {\cal T} \Delta L,
\end{equation}
where $I$ is the $2\times2$ unity matrix, and $\Delta
L\rightarrow0$ is the width of the impurity barrier. This means
that $\Delta\mathbf{\psi}(x)\equiv\psi(x+ \Delta L /2 )-\psi(x-
\Delta L/ 2)=i {\cal T}\mathbf{\psi}(x)\Delta L $, and the matrix
${\cal T}$ is obtained from the equations of motion (\ref{SEOM}),
\begin{equation}
{\cal T}= \left(
\begin{array}{cc} \omega &
 -W \\
W & -\omega \end{array} \right).
\end{equation}
Therefore, we can write the transmission matrix as $M=e^{i
\sigma_{a} \xi^{a}}$, where $\sigma_{a}$ are the Pauli matrices
and $\xi^{1}=0$, $\xi^{2}=-i W \Delta L$, $\xi^{3}=\omega \Delta
L$. Since the backscattering potential $W(x)$ is constant for
$-L/2<x<L/2$, the transmission matrices of vanishingly narrow
impurities commute with each other,
which allows us to find parameters $\xi^a$ for a rectangular
potential with a finite length $L>0$ in the form $\xi^{1}=0$,
$\xi^{2}=-i W L$, and $\xi^{3}=\omega L$. On the other hand,
\begin{equation}
e^{i \sigma_{a} \xi^{a}} = \left( \begin{array}{cc} \cos(\xi)+i
\frac{\xi^{3}}{\xi} \sin(\xi) &
  \frac{i \xi^{1}+\xi^{2}}{\xi} \sin(\xi)\\
\frac{i \xi^{1}-\xi^{2}}{\xi} \sin(\xi) & \cos(\xi)-i
\frac{\xi^{3}}{\xi} \sin(\xi)\nonumber
\end{array}
 \right),
\end{equation}
where $\xi\equiv\sqrt{(\xi^{1})^{2}+(\xi^{2})^{2}+(\xi^{3})^{2}}$.
Therefore, $\xi=k L$ and we obtain Eq.\ (\ref{Mmatrix}).

\end{document}